\documentclass[twocolumn]{openjournal}

\usepackage{xcolor}
\usepackage{amsmath}
\usepackage{textgreek}
\usepackage[utf8]{inputenc}
\usepackage[english]{babel}
\usepackage{soul}

\makeatletter
\expandafter\let\csname longtable*\endcsname\relax
\expandafter\let\csname endlongtable*\endcsname\relax
\makeatother
\usepackage{subcaption}

\makeatletter
\expandafter\let\csname longtable*\endcsname\relax
\expandafter\let\csname endlongtable*\endcsname\relax
\makeatother

\usepackage{hyperref}
\hypersetup{
    unicode, 
    colorlinks=true,
    linkcolor=linkcolor,
    citecolor=linkcolor,
    filecolor=linkcolor,
    urlcolor=linkcolor,
    pdftitle={Inferring Central Galaxies and Halo Masses from Ten Neighbours},
    pdfauthor={Alice Chen, Camille Avestruz, Niayesh Afshordi, and Yen-Ting Lin},
    pdfkeywords={central galaxies, satellite galaxies, halo mass, group finding, galaxy-halo connection, domain adaptation, XGBoost, IllustrisTNG, EAGLE},}
\usepackage{color,colortbl}
\definecolor{linkcolor}{rgb}{0.0,0.3,0.5}
\usepackage{tensind}
\tensordelimiter{?}
\DeclareGraphicsExtensions{.bmp,.png,.jpg,.pdf}
\usepackage{verbatim}
\usepackage[normalem]{ulem}
\usepackage{orcidlink}

\graphicspath{ {./figs/} }

\newcommand{\beq}{\begin{equation}}
\newcommand{\eeq}{\end{equation}}
\newcommand{\bea}{\begin{eqnarray}}
\newcommand{\eea}{\end{eqnarray}}
\newcommand{\beann}{\begin{eqnarray*}}
\newcommand{\eeann}{\end{eqnarray*}}

\newcommand{\xgb}{\texttt{XGBoost}}

\begin{document}

\title{Cross-Dataset Inference of Central-Satellite Classification and Host Halo Masses from Galaxy neighbours with Machine Learning}

\author{Alice Chen\orcidlink{0000-0003-4780-5554}}
\email{ay7chen@uwaterloo.ca}
\affiliation{Department of Physics and Astronomy, University of Waterloo, 200 University Avenue West, Waterloo, Ontario N2L 3G1, Canada}
\affiliation{Waterloo Centre for Astrophysics, University of Waterloo, 200 University Avenue West, Waterloo, Ontario N2L 3G1, Canada}
\affiliation{Perimeter Institute for Theoretical Physics, 31 Caroline Street North, Waterloo, Ontario N2L 2Y5, Canada}

\author{Camille Avestruz\orcidlink{0000-0001-8868-0810}}
\email{cavestruz@umich.edu}
\affiliation{Department of Physics, University of Michigan, Ann Arbor, Michigan 48109, USA}
\affiliation{Leinweber Institute for Theoretical Physics, University of Michigan, Ann Arbor, Michigan 48109, USA}

\author{Niayesh Afshordi\orcidlink{0000-0002-9940-7040}}
\email{nafshordi@pitp.ca}
\affiliation{Department of Physics and Astronomy, University of Waterloo, 200 University Avenue West, Waterloo, Ontario N2L 3G1, Canada}
\affiliation{Waterloo Centre for Astrophysics, University of Waterloo, 200 University Avenue West, Waterloo, Ontario N2L 3G1, Canada}
\affiliation{Perimeter Institute for Theoretical Physics, 31 Caroline Street North, Waterloo, Ontario N2L 2Y5, Canada}

\author{Yen-Ting Lin\orcidlink{0000-0001-7146-4687}}
\email{ytl@asiaa.sinica.edu.tw}
\affiliation{Academia Sinica Institute of Astronomy and Astrophysics, 11F of Astronomy--Mathematics Building, AS/NTU, No. 1, Section 4, Roosevelt Road, Taipei 10617, Taiwan}

\begin{abstract}
The classification of galaxies into centrals and satellites gives insight into galaxy-halo connection.  This, together with halo mass estimates, can be used to connect observed galaxy populations to underlying dark matter structure.  To (1) classify galaxies as centrals or satellites and (2) estimate the host halo mass for centrals, we train \xgb\ models on IllustrisTNG (TNG100) galaxies, using the distances (3D or projected (2+1)D redshift space) and stellar masses of their ten nearest neighbours.  
We then test the model on EAGLE to measure its robustness after applying a simple domain-alignment step that corrects for a relative shift in the stellar-mass distribution between TNG100 and EAGLE.  We further augment the training data by combining feature sets from neighbour samples with varying lower stellar-mass thresholds.
On the TNG100 test set, the central/satellite classifier reaches purity and completeness $=0.97$ ($0.95$) in 3D (projected redshift) space at the threshold where purity=completeness, and the halo-mass regressor achieves $r^2=0.98$ ($0.93$) when evaluated on true centrals and $r^2=0.93$ ($0.93$) when evaluated on inferred centrals. Applied to EAGLE after alignment, the classifier achieves purity/completeness $=0.95$ ($0.95$), while halo-mass inference yields $r^2=0.91$ ($0.87$) for true centrals and $r^2=0.83$ ($0.77$) for inferred centrals; training-data augmentation improves the halo-mass estimation task the most.  Satellites misclassified as centrals underpredict the true masses of the host halos, especially if they are closer to a lower mass halo's central than its own host's central and/or has a significantly lower stellar mass than its own central.  These results show that survey-accessible information such as projected separations, line-of-sight distances, and stellar masses - combined with simulation-trained models and domain alignment - can jointly enable reliable central identification and host halo mass estimation to an $r^2 \geq 0.85$.

\end{abstract}

\begin{keywords}
    {galaxies, methods:data analysis, dark matter halo}
\end{keywords}

\maketitle

\section{Introduction}
In the $\Lambda$CDM model, structure formation occurs hierarchically.  Galaxies form within dark matter potential wells, where baryonic matter is gravitationally pulled into dark matter overdensities and subsequently cools and collapses, while the surrounding dark matter structure evolves into a virialized halo \citep{1991ApJ...379...52W, Navarro:1995iw}.  As a result, galaxies merge and evolve in tandem with their host halos.  This eventually becomes a structure where multiple galaxies, known as satellites, occupy the halo of the gravitationally dominant galaxy of that halo, known as a central \citep{Peacock:2000qk, Cooray:2002dia, Berlind:2002rn, 10.1093/mnras/275.1.56, Wechsler:2018pic}.  Because central galaxies formed and evolved with their host halos, their properties should have the stronger correlations with the host halo mass \citep{2016ApJ...819..119L, Bilicki:2021hgn, 2023OJAp....6E..37B,2025arXiv251019926D}.  Being able to classify galaxies as centrals or satellites can then be useful when trying to determine environment \citep[e.g., see][for a recent discussion]{Bowden:2025djs} and mass of the host halo, the latter which primarily consists of dark matter.  The host halo mass can then be used to construct the halo mass function and density profiles \citep{Cooray:2002dia, Asgari:2023mej, Tinker:2008ff}, which are sensitive to cosmology and can be a probe of cosmological parameters.

In observations it is often unclear how to distinguish a satellite galaxy from a central, as the underlying host halo is made mostly of dark matter and thus unobservable.  Central and satellite galaxies often have different properties, as satellites form outside the present central's host halo and fall into the halo at a later time.  Thus, they are subject to the ram pressure, harassment, and tidal forces of the host halo and its central, whereas the central galaxy would be significantly less subject to these processes.  Consequently, satellite galaxies would have different formation histories compared to those of central galaxies.  

\begin{figure*}[htp!]
    \centering
    \subfloat[]{
    \includegraphics[width=0.95\linewidth]{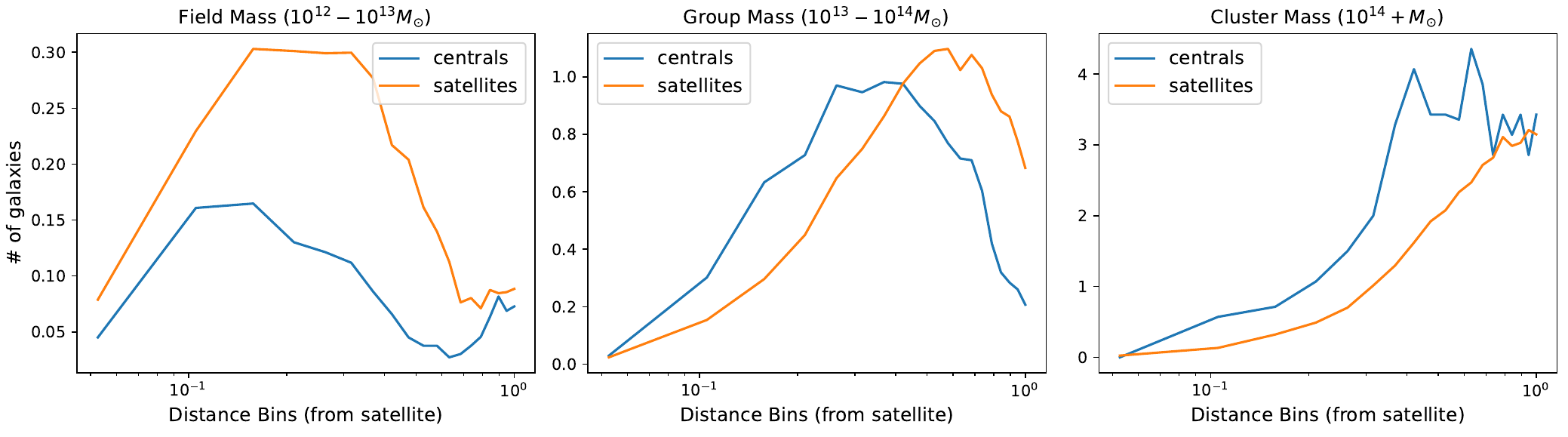} }
    \vspace{0.1em}
    \subfloat[]{
    \includegraphics[width=0.95\textwidth]{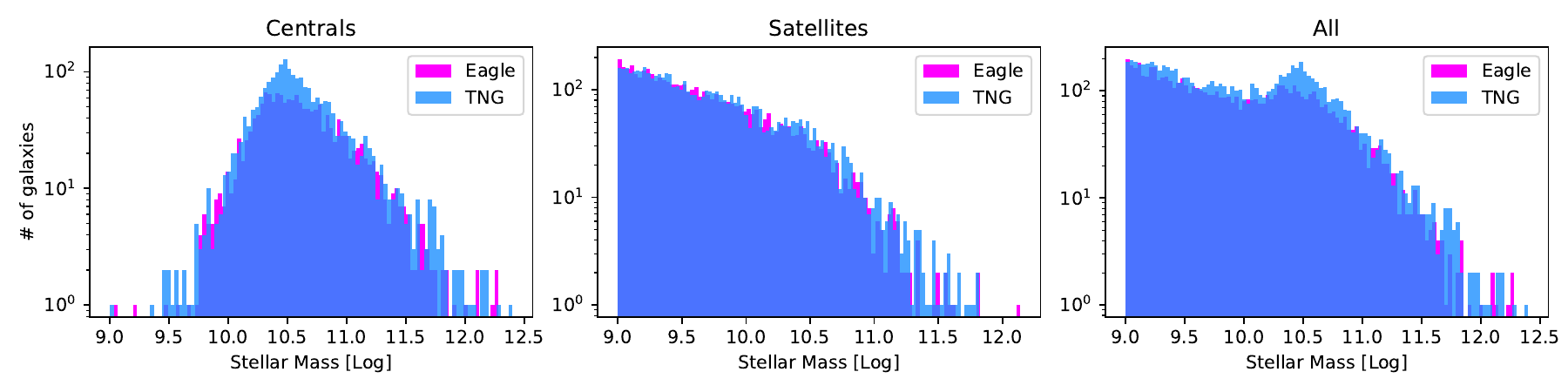}}
    \caption{a) (Top) Total number of galaxies per radial bin, in 3D distance bins.  The radial profiles are stacked for either centrals or satellites in the same host halo mass bin: $\log M_{host,vir}/M_\odot\in[12,13),[13,14),[14,14.5]$. The galaxy radial distribution around centrals is distinct from those of satellites.  It should be noted that for cluster sized halos, in the right most panel, $M_{halo}\geq 10^{14}M_{\odot}$, TNG100 only had 17 of these objects.  b) (Bottom) The stellar mass distribution of galaxies in central galaxies vs satellites, and all galaxies (satellites + centrals).  It can be seen that centrals have a peak at around $10^{10}-10^{11}M_{\ast}$, while satellite with higher stellar masses continuously decrease.  This shows a stellar mass gap between centrals and satellites, which leads to a magnitude gap in observations.}
    \label{fig:rad_1Mpc_neighbours}
\end{figure*}

Here, we aim to develop a method to identify central galaxies, building upon the results of \citet{Oguri:2014dqa}, where the mass of the underlying host halo of a given galaxy can be determined based on its number of nearest neighbours within some fixed cylindrical radii; the optimal cylindrical radius for this is found to be 0.5Mpc/h.  From their results and Figure \ref{fig:rad_1Mpc_neighbours}, it can be seen that spatial clustering of neighbours around each galaxy can be a method used to distinguish between centrals and satellites, as their neighbouring galaxy distribution curves differ from each other.  This difference in neighbouring galaxy distribution can be used to determine whether a galaxy is a central or a satellite.  In most previous works, centrals were often identified using stellar mass and luminosity \citep[e.g see,][for a reference for observational analyses]{SDSS:2004oes, 2005ApJ...633..791Z}.  However, the brightest galaxy is not necessarily the central, as seen in \citet{2011MNRAS.410..417S}.
Consequently, we add stellar mass into our classification training data alongside the neighbouring distances, since that is observationally available, but we do not make prior assumptions about the stellar mass or luminosity of the galaxy and its classification.  The goal is utilising both to yield more accurate central classification than what has been done before.

In this work, we used data from the Illustris-TNG simulations \citep{illustris_tng, Nelson:2018uso} to train \xgb\ models to classify galaxies as centrals and satellites, and to perform host halo mass estimations for identified centrals.  In section \ref{sec:methods}, we describe the simulation data used, including the EAGLE simulation \citep{Schaye:2014tpa}, which we use as a hold-out test dataset for model evaluation.  We also briefly describe the \xgb\ machine learning algorithm, our input feature selection, and experiments for model training.  In section \ref{sec:results}, we show results of classification and host halo mass estimation from feature subsets that include neighbour galaxy distances and stellar masses.  We also present results when the method is applied to the EAGLE simulation, whose galaxy formation model is distinct from that implemented to produce TNG.  In section \ref{sec:discussion}, we provide interpretations of our results in the context of other work, describe potential benefits of our approach, in addition to some limitations.  Finally, in section \ref{sec:conclusion}, we summarize this paper and outline some future prospects.

\section{Data and Methodology}\label{sec:methods}

\subsection{Simulation Data: Illustris-TNG and EAGLE}\label{sec:TNGvEAGLE}
Our analysis uses data from the publicly available Illustris-TNG \citep{Nelson:2018uso, illustris_tng} and EAGLE cosmological simulation suites \citep{Schaye:2014tpa}.  Specifically, we focus on the TNG100 simulation and EAGLE RefL0100N1504.  These simulations provide comparable volumes (110 Mpc vs. 100 Mpc) and mass resolutions, with similar numbers of gas and dark matter particles.  While TNG100 adopts a Planck 2015 cosmology \citep{Planck:2015fie} and EAGLE adopts a Planck 2013 cosmology \citep{Planck:2013pxb}, both simulations produce similar galaxy stellar mass and host halo mass distributions.

In both public catalogs, we identify the FoF halos and SUBFIND subhalos.  In the TNG group catalog, the relevant fields are \texttt{GroupMass}, corresponding to the total FoF group mass, \texttt{GroupFirstSub} for identifying the central, \texttt{SubhaloFlag} for removal of non-cosmological fragments \footnote{halos deemed by the TNG algorithm as being a numerical artifact and thus not a physical object}, \texttt{SubhaloPos} for co-moving position in ${\rm ckpc}\,h^{-1}$, and the stellar component of \texttt{SubhaloMassType}.  In the analogous EAGLE catalogue, \texttt{GroupMass} corresponds to the total Friends-of-Friends (FOF) mass of this halo, \texttt{SubGroupNumber}, \texttt{Flag}, Cartesian coordinates, velocities, and \texttt{MassType\_Star} \citep{McAlpine:2016eagle, Nelson:2018uso}.

The analyses we report here use the snapshots at $z=0$ ($a=1$).  The lengths used to calculate distance are in units of $h^{-1}Mpc$, while the mass is in units of $h^{-1} M_\odot$.  The EAGLE catalog stores positions in $Mpc$ and masses in $h^{-1} M_\odot$; for distance calculations its positions were multiplied by $h_{\rm EAGLE}\simeq0.678$, so the 100-Mpc box spans approximately $67.8\,h^{-1}{\rm Mpc}$.  The two simulations were therefore compared using distances in $h^{-1}{\rm Mpc}$ and masses in $h^{-1} M_\odot$.

To ensure that our sample consists of well-resolved galaxies that would be observable in real surveys, we impose a stellar mass cut of $M_* \geq 10^9M_\odot$, since galaxies below this cut consist of less than 1000 stellar mass particles and likely do not behave the same way physical galaxies in observations do.  Galaxies flagged by the simulations as unphysical are removed from the sample as well.  We additionally subselect galaxies to have host halo masses of $M_{\rm halo} \geq 10^{12}M_\odot$.  This is to preserve a group halo environment for the galaxies to reside in \citep{Donnari:2020xof}, so that each halo would have at least $\geq 1$ satellite \citep{Kennedy:2026nlx}.  The galaxies themselves were also further subselected to follow the standard stellar-to-halo mass relation - objects with non-standard stellar mass fractions ($M_{\ast}/M_{total} > 20 \%$) are excluded.  This is because we cannot distinguish between numerical resolution effects or physical formation anomalies for galaxies having an irregular stellar-to-halo mass, so we exclude these galaxies to minimize the possibility of numerical artifacts.

Our fiducial sample adopts a stellar mass cut of $10^9 M_\odot$.  However, observational galaxy catalogues often employ different stellar mass limits due to survey depth and resolution limitations.  To investigate the robustness of our methods to these selection effects, we additionally combine samples constructed from varying the lower limit of stellar mass.  Throughout this work, true centrals and satellites are assigned using the subhalo classifications given by the simulation.  Central galaxies are defined as the primary galaxy residing at the centre of a host halo, while satellites are all other remaining galaxies gravitationally bound to the host halo.  We use the galaxies' stellar masses and neighbouring distances to get as close to the actual galaxy classification as possible.  The lower panel of Figure \ref{fig:rad_1Mpc_neighbours} compares the stellar mass distributions of the selected TNG100 and EAGLE samples. Due to differences in subgrid physics and stellar formation models, there are discrepancies between the stellar mass functions of the two simulations.  

\begin{figure}[htp!]
    \centering
    \includegraphics[width=0.95\linewidth]{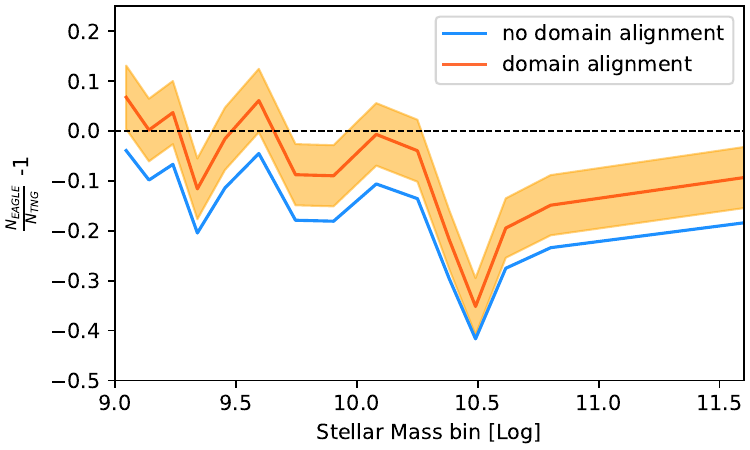}
    \caption{Stellar mass ratio differences between TNG and EAGLE.  Due to differences in galaxy formation models, the EAGLE stellar mass function is systematically lower except at the very high stellar mass ends, likely because EAGLE uses a higher $\sigma_8$. Dividing EAGLE stellar masses by a factor of 0.91 minimizes the difference in distributions using the KS test. The shaded area shows the 1-$\sigma$ Poisson error.}
    \label{fig:smf_comparison_and_correction}
\end{figure}

To account for this, we applied an empirical stellar mass domain alignment to TNG100 in order to better match the stellar mass distribution of EAGLE.  Figure \ref{fig:smf_comparison_and_correction} shows the ratio of original stellar mass functions from before and after applying the domain alignment.  This prefactor was determined by minimizing the difference between the cumulative stellar mass distributions of the two simulations. We constructed empirical cumulative distribution functions (CDFs) of the stellar masses and evaluated their agreement using a Kolmogorov-Smirnov (KS) statistic, then selected an optimal prefactor as the value that minimized the KS distance between the two CDFs. This procedure is intended to emulate the calibration process that would be required when applying our methodology to observational catalogues, where differences in stellar mass estimates between surveys are common and often require empirical corrections before samples can be combined.

When training and testing on TNG, we use a 75:25 random galaxy split: 75$\%$ of the sample become our training set, while the remaining $25\%$ is the testing set.  We note that some neighbours of training set galaxies may also be neighbours of the test set galaxies. However, it is only the neighbour stellar mass that is redundant, and the neighbour distances capturing the actual environment are distinct. So, we do not consider this to be a source of leakage between the training and testing sets.  When training on TNG and testing on EAGLE, we take the entire TNG set of selected galaxies for training, and the set of selected EAGLE galaxies for testing.  In this case, there are no overlapping galaxies or host halos at all in either training or testing.

\subsection{XGBoost Classification and Halo Mass Estimation}

For our analysis, we used the Extreme Gradient Boosting (XGBoost) algorithm \citep{2016arXiv160302754C}. XGBoost is an ensemble machine learning method based on gradient-boosted decision trees and is widely used due to its computational efficiency, robustness to complex nonlinear relationships, and strong predictive performance across a broad range of applications.  Given these strengths, and the fact that it would take a multi-layer neural network with heavier computational efforts to obtain similar results (see Table \ref{tab:scores}, where \xgb\ had the best performance), we decided to use XGBoost as our main classifying and regression tool here.

XGBoost was trained to distinguish between central and satellite galaxies using the distances and stellar masses from the sub-samples described above.  For host halo mass estimation, the algorithm is trained as a regression model that predicts the mass of a galaxy's host halo from the same set of input features.  Hyperparameters were optimized using cross-validation on the training set.  We also tested logistic regression and a two-layer neural network as alternative classification models, but they consistently produced lower classification and mass estimation accuracy compared to XGBoost (see Table \ref{tab:scores}).

\subsection{Feature Selection}

We explored subsets of input features motivated by both observational accessibility and theoretical correlation with the galaxy type and host halo mass.  These included nearest-neighbour distances, neighbouring galaxy stellar masses, and the stellar mass of the target galaxy itself.

To assess the robustness of our results to observational selection effects, feature selection was repeated for samples defined by different stellar mass cuts.  The stellar mass cut was also included as an additional training parameter when the training set combined varying stellar mass cuts together.  We found that models trained using multiple stellar mass cut samples consistently outperformed those trained using a single fiducial selection, likely because there are more galaxies to train on.  This finding indicates that information from galaxies spanning a range of varying stellar mass cuts provides additional constraints on halo mass and galaxy type.

We present the results for two sets of training distances: 3D spatial distance in $(x,y,z)$ and (2+1)D redshift space distance.  The redshift space distance is calculated as the 2D projected distance in $(x,y)$, with an additional redshift training component estimated as $H_0\Delta z + \Delta V_z$, where $\Delta V_z$ is the difference in the galaxies' z-direction velocity, and $H_0$ is the Hubble constant, and $\Delta z$ is the different in the galaxies z-direction distance.  We further constrained the $z-$axis distance to be $\pm 2.0 Mpc/h$ to prevent finding ``neighbours" that seem close in the 2D $xy$ plane, but are actually very far from each other in the $z-$direction.  This is a closer proxy to what we would observe in observations, since objects are observed in redshift space, not 3D real space.

We also experimented with training based on galaxy properties that were quantile transformed \citep{2005math.....11508D}. This procedure maps each feature to a target distribution based on its cumulative rank ordering, reducing the influence of extreme outliers and highly skewed distributions.  However, we found that quantile transformation of the data did not yield an improvement in performance results, so our final results do not include quantile transformation.

\section{Results}\label{sec:results}
Here, we present the performance of our central/satellite galaxy classification and halo mass prediction algorithms, based on the neighbouring distance and stellar mass information.  

From Figure \ref{fig:rad_1Mpc_neighbours}, we can see that the neighbouring galaxy distribution differs around central galaxies compared to satellites.  Making use of the 10 closest neighbouring distances, the galaxies' stellar masses, and XGBoost, we were able to classify galaxies in the TNG100 and EAGLE simulations to an accuracy of $>90\%$.  The results are shown in subsection \ref{sec:results_cen_sat}. 

Since halos and their central galaxies evolve together, there should be a correlation between the halo mass and the neighbouring galaxy distances of its central galaxy; halo mass would affect satellite density, infall, and thus inner galaxy distribution.  Using the centrals we classified in subsection \ref{sec:results_cen_sat}, we also infer the mass of their host halos, again from the TNG100 and EAGLE simulations.  The results are presented in subsections \ref{sec:results_tng_mass} and \ref{sec:results_eagle_mass}.

\subsection{Central vs. Satellite Galaxy Classification}
\label{sec:results_cen_sat}

In this subsection, we show the results of central/satellite classification that are trained on the TNG simulation only, then tested on both an independent subset of TNG data, and on a separate simulation, EAGLE.  This is to see the robustness of our model across different simulations with different subgrid physics, such as star formation and AGN feedback models.  We also have a fiducial mass cut of $M_{\ast}\geq 10^9 M_{\odot}$, and a mixture of galaxies from different sub-selected mass cuts ($M_{\ast}\geq 10^9 M_{\odot}, 10^{10} M_{\odot}, 10^{11} M_{\odot}$).  The differing mass cut for the mixed sample was also an input parameter used in the training process.

\begin{figure*}[ht]
    \begin{subfigure}[t]{0.48\linewidth}
        \centering  
        \includegraphics[width=0.95\linewidth]{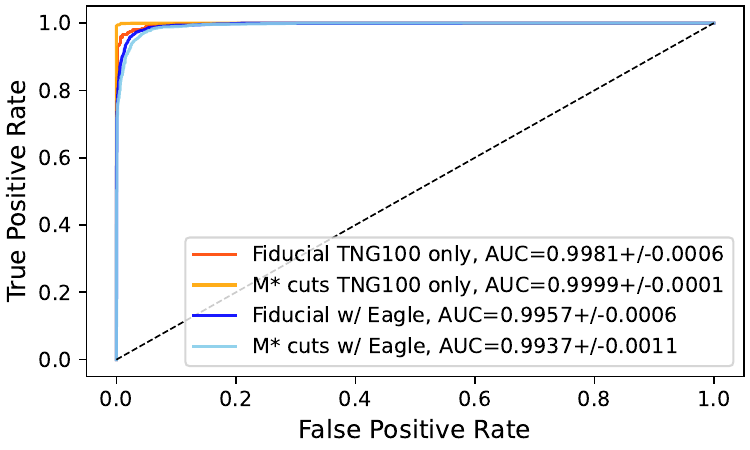}
        \caption{Receiving operating characteristic (ROC) curve (left) for the classification of central galaxies.  We show results for an \xgb\ model trained with TNG100 data to identify centrals based on neighbour galaxy distances selected on a fixed $M_*$ cut, and also an \xgb\ model trained to identify centrals based on neighbour galaxy distances selected from varied $M_*$ cuts.  Varying the neighbour selection in model training improves classification when evaluated on hold-out test data from the TNG100, but not when applied to the EAGLE dataset. }
        \label{fig:3d_roc}
    \end{subfigure}
    \begin{subfigure}[t]{0.48\linewidth}
        \centering        \includegraphics[width=0.95\linewidth]{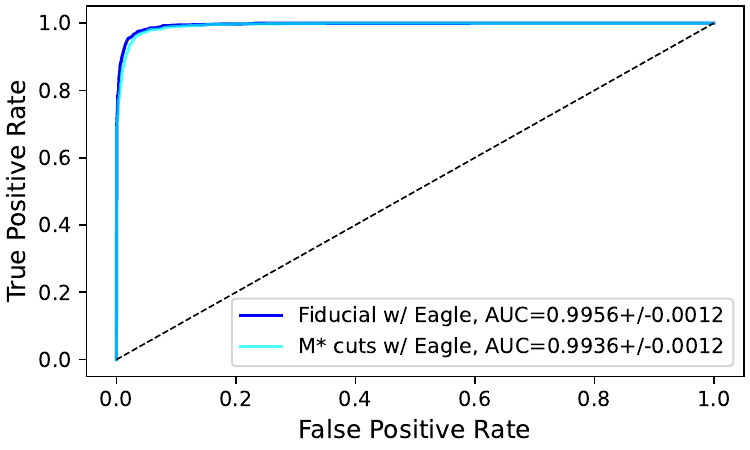}
        \caption{Same as the the plot on the left, except now using the 2D projection space.  We include the results of training on TNG and testing on EAGLE, since that is a comparatively better proxy for real space observations, if we train on simulation data.}
        \label{fig:2d_roc}
    \end{subfigure}
    \centering
    \caption{Galaxy classification results.  ``Fiducial'' uses neighbours with $M_*\geq10^9\,M_\odot$; ``Different $M_*$'' combines neighbour selections at $10^9$, $10^{10}$, and $10^{11}\,M_\odot$}
    \label{fig:roc_summary}
\end{figure*}

\begin{figure*}[ht]
    \begin{subfigure}[t]{0.46\linewidth}
        \centering
        \includegraphics[width=0.95\linewidth]{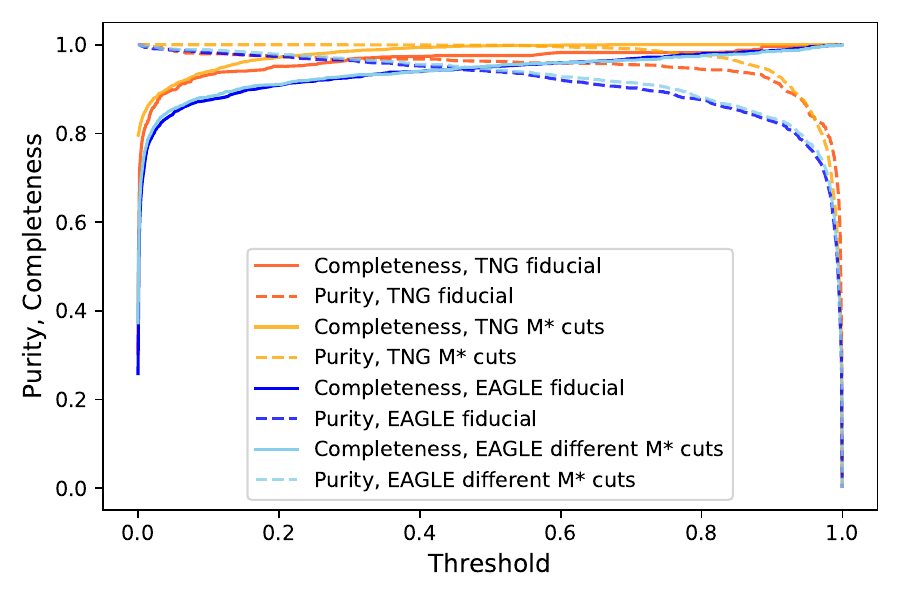}
        \caption{Purity (dashed lines) and completeness (solid) curve of our \xgb\ model predictions of centrals using the full 3D neighbouring distances.}
        \label{fig:3d_thresh}
    \end{subfigure}
    \begin{subfigure}[t]{0.46\linewidth}
        \centering
        \includegraphics[width=0.95\linewidth]{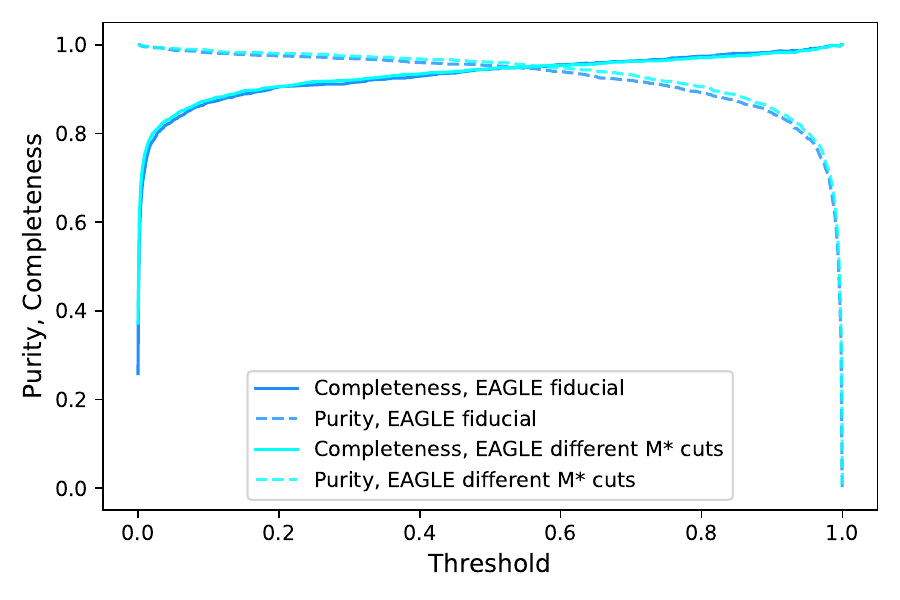}
        \caption{The purity-completeness curves using the 2D projection space distance.}
        \label{fig:2d_thresh}
    \end{subfigure}
    \caption{Purity-completeness curves.  Orange lines show metrics that correspond to a model trained and evaluated on TNG data, with a fiducial fixed $M_*$ cut having a dark shade and varying $M_*$ cuts having a lighter shade.  Blue lines show metrics that correspond to a model trained on TNG data and tested on EAGLE, again with a dark shade representing a fiducial fixed $M_*$ cut and a lighter shade representing varying $M_*$ cuts}
    \label{fig:cp_summary}
\end{figure*}

\begin{figure*}[htp!]
    \begin{subfigure}[t]{0.48\linewidth}
        \centering        \includegraphics[width=0.95\linewidth]{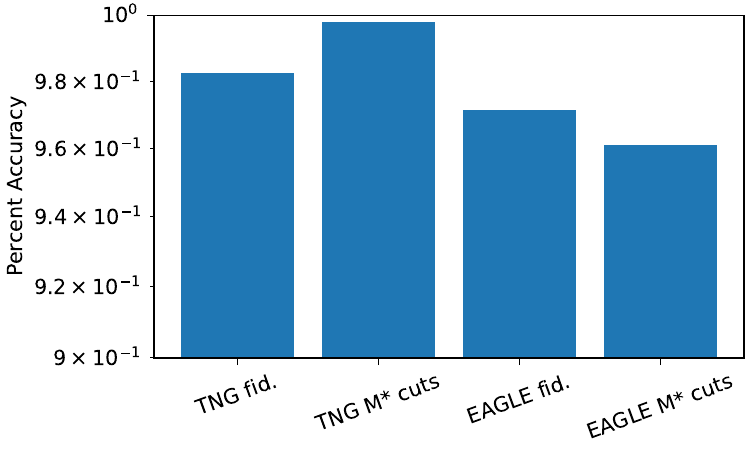}
        \caption{Classification done using 3D distances.}
    \end{subfigure}
    \begin{subfigure}[t]{0.48\linewidth}
        \centering        \includegraphics[width=0.95\linewidth]{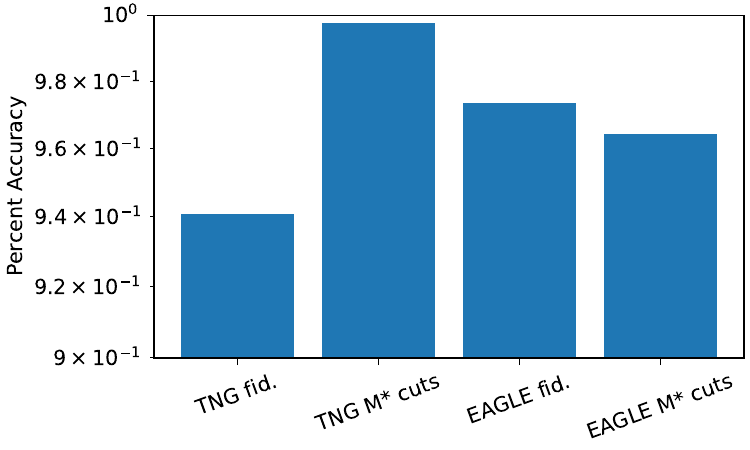}
        \caption{Classification done using (2+1)D projected distances.}
    \end{subfigure}
    \centering
    \caption{Accuracy of the central and satellite classification.}
    \label{fig:bar_summary}
\end{figure*}

\begin{figure*}[t!]
    \begin{subfigure}[t]{0.48\linewidth}
        \centering        \includegraphics[width=0.95\linewidth]{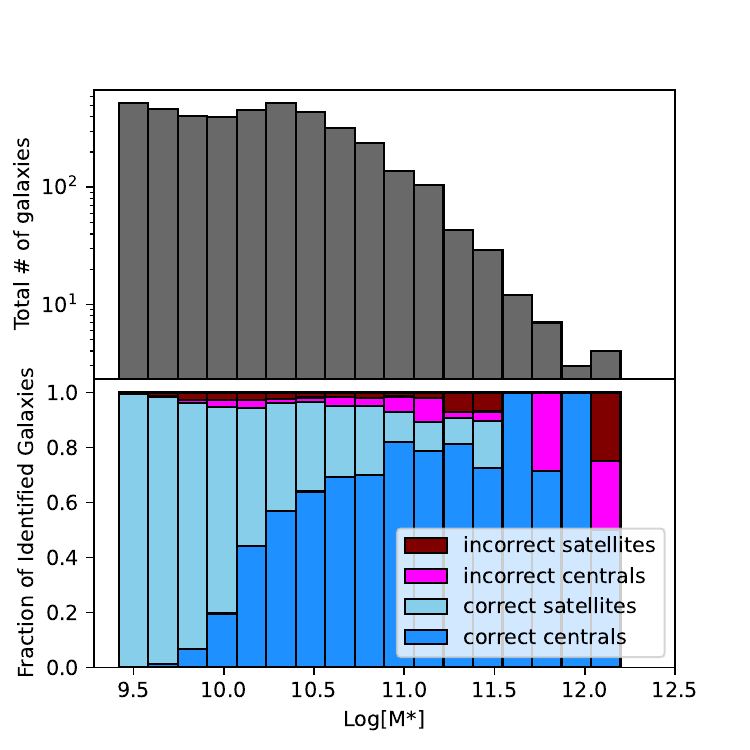}
        \caption{Probability accuracy of (2+1)D projected distances.}
    \end{subfigure}
    \begin{subfigure}[t]{0.48\linewidth}
        \centering        \includegraphics[width=0.95\linewidth]{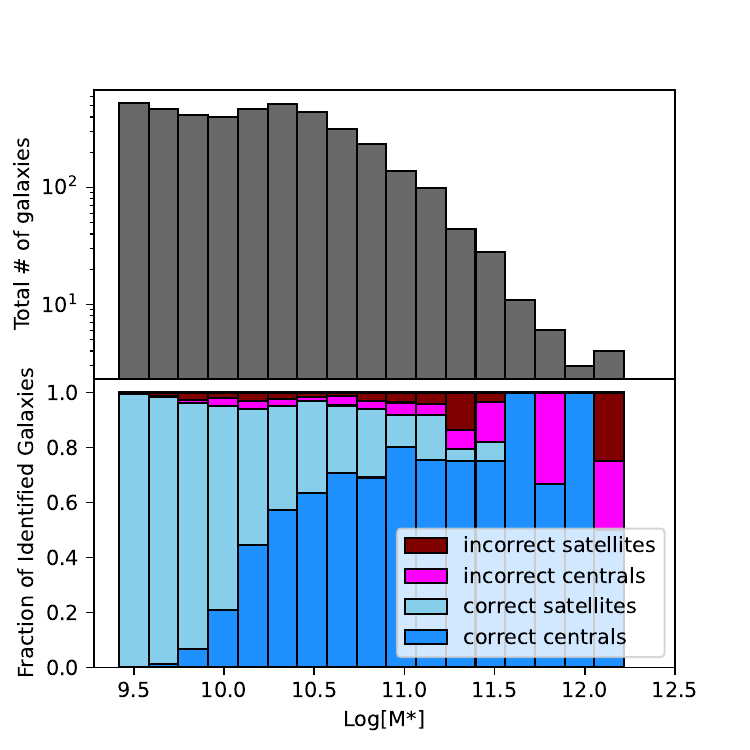}
        \caption{Probability accuracy of full 3D distances.}
    \end{subfigure}
    \centering
    \caption{Bar plot of the probability of obtaining true/false positives and negatives of the central and satellite classification.}
    \label{fig:correctness_summary}
\end{figure*}

From the receiving operating characteristic (ROC) curves in Figure \ref{fig:roc_summary}, it can be seen that using the distances, mass-cuts, and stellar masses of the galaxies, the classification algorithm works quite well - all ROC curves have an area under curve (AUC) of $>0.99$ and the overall probabilities are predicted with a success rate of $\geq 90\%$ (see Figures \ref{fig:bar_summary} and \ref{fig:correctness_summary}), for testing on TNG and EAGLE.  Training and testing on the same dataset (TNG) does yield a slightly better result than training and testing on a different dataset, due to the galaxy formation mechanisms for the testing and training sets being identical, but even across the different simulations, the AUCs are still very close to 1, which indicates a very good classification.  These results are further bolstered by Figure \ref{fig:cp_summary} and Table \ref{tab:purity_values}, which both show purity and completeness values of $\geq 94\%$ at the threshold where the two are equal, across all training and testing sets. Quantile transformation of the data also did not yield an improvement in the results (see \ref{tab:scores} in the Appendix).

\begin{table*}[]
    \centering
    \begin{tabular}{|c|c|c|c|c|c|c|}
        \hline
        Distance & Test sample & Selection & $P=C$ value & Threshold & AUC & Avg. Precision \\
        \hline
        3D & TNG & One $M_{\ast}$-cut & 0.9673 & 0.3017 & 0.9981 & 0.9958 \\
        \hline 
        3D & TNG & Different $M_{\ast}$-cut & 0.9983 & 0.5385 & 0.9999 & 0.9999 \\
        \hline 
        3D & EAGLE & One $M_{\ast}$-cut & 0.9451 & 0.4547 & 0.9957 & 0.9884 \\
        \hline 
        3D & EAGLE & Different $M_{\ast}$-cut & 0.9489 & 0.4688 & 0.9937 & 0.9895 \\ 
        \hline
        (2+1)D & TNG & One $M_{\ast}$-cut & 0.9611 & 0.01073 & 0.9804 & 0.9910 \\
        \hline 
        (2+1)D & TNG & Different $M_{\ast}$-cut & 0.9974 & 0.3358 & 0.9998 & 0.9999 \\
        \hline 
        (2+1)D & EAGLE & One $M_{\ast}$-cut & 0.9481 & 0.5363 & 0.9956 & 0.9881 \\
        \hline 
        (2+1)D & EAGLE & Different $M_{\ast}$-cut & 0.9485 & 0.5755 & 0.9936 & 0.9897 \\
        \hline 
    \end{tabular}
    \caption{\textbf{Central/Satellite Classification Metrics: }Common purity and completeness, $P=C$, at the crossover threshold.  Again, here, ``Fiducial'' uses neighbours with $M_*\geq10^9\,M_\odot$; ``Different $M_*$'' combines neighbour selections at $10^9$, $10^{10}$, and $10^{11}\,M_\odot$.  Thresholds refer to the uncalibrated \xgb\ score.}
    \label{tab:purity_values}
\end{table*}

Figure \ref{fig:bar_summary} show the overall accuracy rate in predicting centrals and satellites; a more detailed version - in both 3D and (2+1)D - of this is in Figure \ref{fig:correctness_summary}.  We can see the percentage of galaxies identified correctly as centrals and satellites, as well as false negatives/positives in each category.  This is comparable to some results in \citet{Bowden:2025djs}, although we have much more satellite galaxies than centrals in our sample in comparison.

\subsection{Host Halo Mass Predictions}

With the galaxy central/satellite classifier working well with high purity scores and very high AUCs for their ROC curves, we take our classified centrals and try to predict their host halo mass.  We start by assuming that we know the true centrals in the simulation, in order to assess the accuracy of neighbourhood-based mass estimation.  We then extend the test to the centrals guessed by XGBoost, to allow for additional confusion introduced by satellite contamination of the sample.  We discuss the results in subsections \ref{sec:results_tng_mass} and \ref{sec:results_eagle_mass} below.

\subsubsection{Predicting host halo masses from TNG}\label{sec:results_tng_mass}

The mass of the host halo depends on its formation history, including mergers, which also affects the distribution of satellite galaxies.  Therefore, there should be a correlation between the host halo mass and the distances of the closest neighbours to its central galaxy.  Because stellar mass is also tightly correlated with the halo mass \citep{Girelli:2020goz}, we include both the stellar mass of the central galaxy and that of its neighbours as additional training parameters.

\begin{figure*}[htp!]
    \centering
    \includegraphics[width=0.95\linewidth]{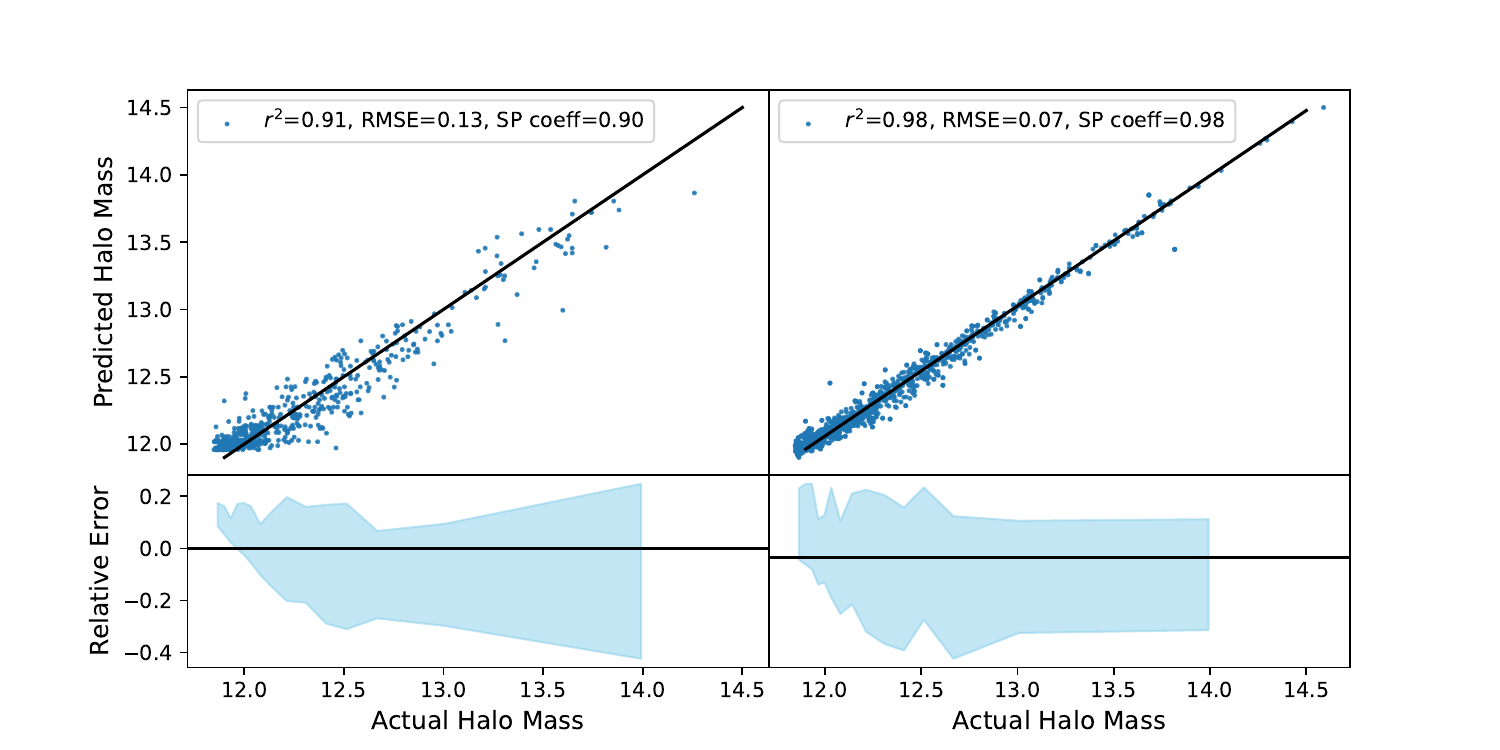}
    \caption{TNG-trained and TNG-tested halo-mass regression for true centrals using 3D separations.  The left panel uses the fiducial $M_*\geq10^9\,M_\odot$ neighbour sample; the right combines the three neighbour thresholds.  The lower panels show the plotted fractional residual statistic.  RMSE is the root mean squared error and SP is the Spearman correlation coefficient.}
    \label{fig:bestmasspred_tng_on_tng}
\end{figure*}

To mimic the testing conditions on potential observations as closely as possible, we train on known centrals, then test on both known and predicted centrals since in observations, we will only have predicted centrals.  We start with training and testing on TNG only to see how well the results generally perform.  From Figure \ref{fig:bestmasspred_tng_on_tng}, it can be seen that both the true and predicted centrals are able to predict their host halo masses reasonably well, with $r^2 \geq 0.9$.

\begin{table*}[ht]
    \centering
    \begin{tabular}{|c|c|c|c|c|c|}
    \hline 
   Distance & Centrals & Neighbour selection & $M_{\ast}$ $r^2$ & RMSE [dex] & $\rho_{\rm S}$ \\  \hline    
    3D & Known & One $M_{\ast}$-cut & 0.91 & 0.13 & 0.90 \\  \hline
    3D & Known & Different $M_{\ast}$-cuts & 0.98 & 0.07 & 0.98 \\  \hline
    3D & Guessed & One $M_{\ast}$-cut & 0.86 & 0.19 & 0.87 \\  \hline
    3D & Guessed & Different $M_{\ast}$-cuts & 0.93 & 0.13 & 0.93 \\  \hline
    (2+1)D & Known & One $M_{\ast}$-cuts & 0.90 & 0.13 & 0.90 \\  \hline
    (2+1)D & Known & Different $M_{\ast}$-cuts & 0.93 & 0.13 & 0.93  \\  \hline
    (2+1)D & Guessed & One $M_{\ast}$-cuts & 0.85 & 0.21 & 0.92  \\  \hline
    (2+1)D & Guessed & Different $M_{\ast}$-cuts & 0.93 & 0.12 & 0.93  \\  \hline
    \end{tabular}
    \caption{\textbf{Same Dataset Host Halo Mass Inference}: Summary comparison of main results for host halo mass predictions on TNG100 for both training and testing.  There are different metrics for the host halo mass regression task with different \xgb\ models.  The model is trained on a subset of TNG100, then tested on a different subset.  The evaluated galaxies are either ``known" true centrals, or predicted centrals from a model trained on TNG data with neighbours selected with different $M_*$ cuts.
    Features include distances to 10 nearest neighbours and the stellar mass of the central (either known true central or predicted central during the classification task) and stellar masses of the neighbours.}
    \label{tab:tng_mass_metrics}
\end{table*}

\subsubsection{Predicting host halo masses in EAGLE from TNG trained models}\label{sec:results_eagle_mass}

The overall goal of this analysis is to be able to predict host halo masses for real galaxies in observational data.  The galaxies from observational catalogues will likely have different star formation mechanisms and thus different stellar mass distributions than in simulations.  As a result, to mimic these unknown underlying differences, we tested on another simulation, EAGLE, without training on it at all; EAGLE has different galaxy and star formation mechanisms than TNG, so it is a proxy for what we would encounter in nature. Similarly to what we did for TNG above, we train on known centrals and test on both known and predicted centrals from EAGLE.

\begin{figure*}[htp!]
    \includegraphics[width=0.95\linewidth]{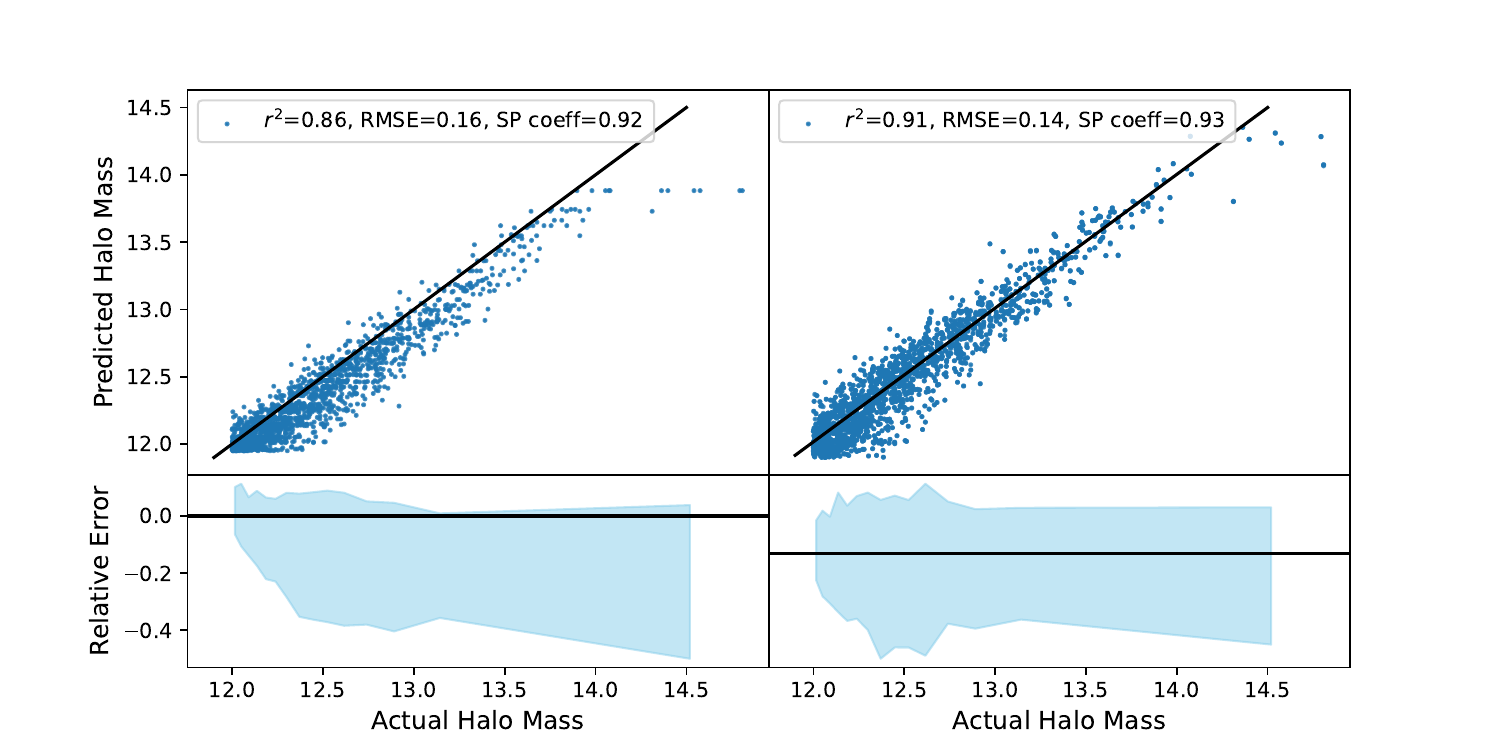}
    \caption{TNG-trained halo-mass regressors evaluated on true EAGLE centrals after stellar-mass CDF alignment, using 3D separations.  The left panel uses the fiducial neighbour threshold and the right combines the three thresholds.  Lower panels show the plotted fractional residual statistic.}
    \label{fig:tng_eagle_mass_pred}
\end{figure*}

The results show that even when trained and tested on different simulations, the predictions remain fairly robust after we correct the stellar masses by a multiplicative factor (see Section \ref{sec:TNGvEAGLE} and Figure \ref{fig:smf_comparison_and_correction}).  This is shown in Table \ref{tab:mass_metrics}.  The $r^2$ values are 0.86 for the fiducial mass cut and 0.91 for combined varying mass cuts when switching the testing to an entirely new simulation, EAGLE - see Figure \ref{fig:tng_eagle_mass_pred}.  Although this is not as good as training and testing solely on TNG, it can be seen here that testing and training on completely different simulations can still yield fairly robust results.

\begin{figure*}[htp!]
    \centering
    \includegraphics[width=0.9\linewidth]{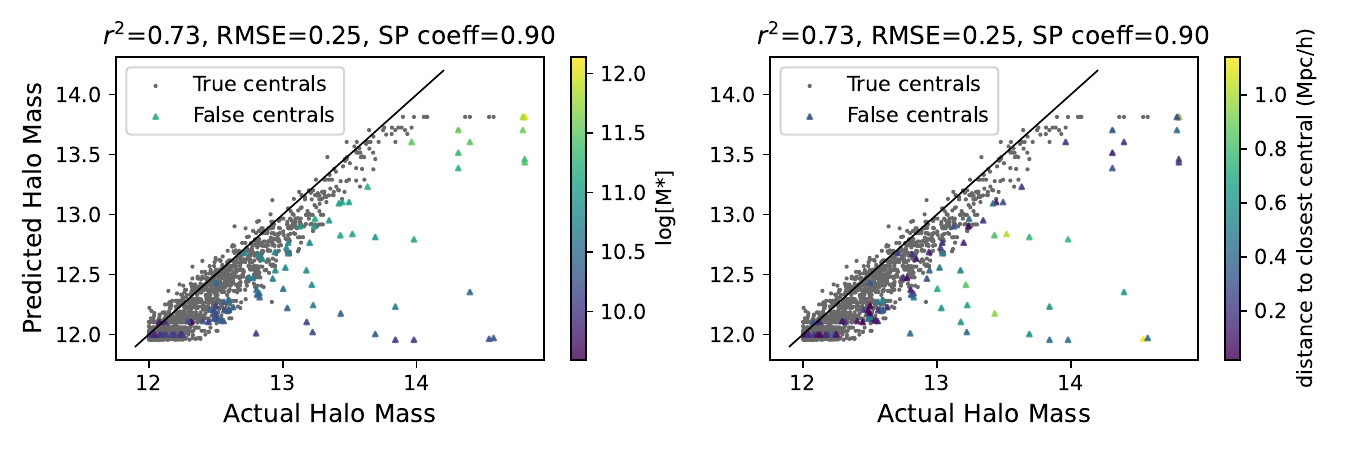}
        \caption{Fiducial mass cut of $10^9 M_{\odot}+$ of the predicted centrals, used to predict their host halos in full 3D space, with the incorrectly identified centrals (actually satellites) coloured according to their stellar mass ratio and 3D distance to its closest central.  Note that the closest central is not necessarily the central of the satellite's host.}
    \label{fig:3d_guess_predict}
\end{figure*}

\begin{figure}
    \centering
    \includegraphics[width=0.9\linewidth]{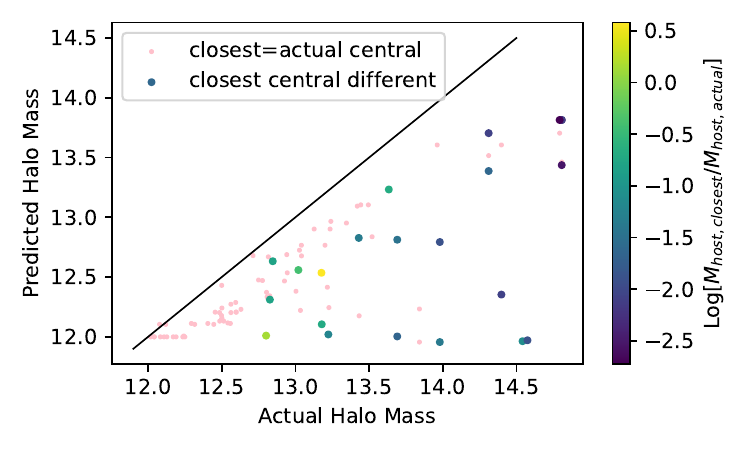}
    \caption{Incorrectly guessed centrals and their host masses in 3D space.  It can be seen that misclassified galaxies closer to another halo's central than their own host halo produces the largest discrepancy between predicted and actual host mass.}
    \label{fig:3d_incorrect}
\end{figure}

\begin{figure}
    \centering
    \includegraphics[width=0.9\linewidth]{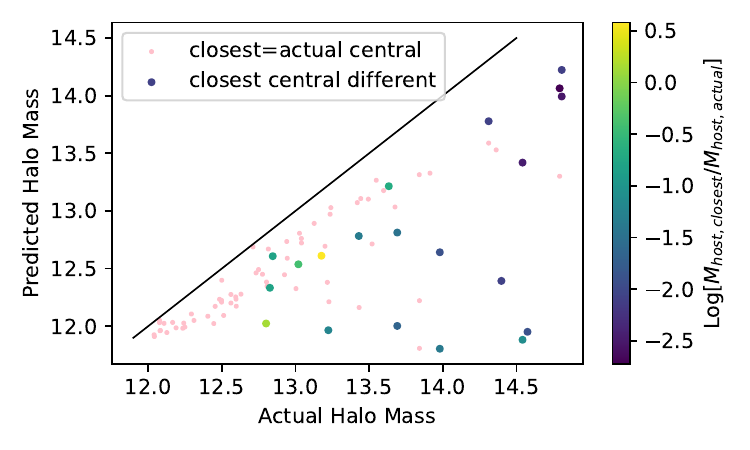}
    \caption{Incorrectly guessed centrals and their host masses in (2+1)D space.  Consistent with the 3D case, it can be seen that misclassified galaxies closer to another halo's central than their own host halo produces the largest discrepancy between predicted and actual host mass.}
    \label{fig:2d_incorrect}
\end{figure}

In addition, we tested our halo mass predictions on guessed centrals instead of true centrals; the results of this are seen in Figure \ref{fig:3d_guess_predict}.  It can be seen here that the largest deviations in predicted host halo mass come from incorrectly classified galaxies, specifically satellites misclassified as centrals.  From the misclassified galaxies in Figures \ref{fig:3d_incorrect} and \ref{fig:2d_incorrect}, we can further see that satellites with higher stellar masses or close to other centrals are more likely to be misclassified.

\subsubsection{Redshift Space Results}
The distances to other galaxies in real observations are most commonly measured in redshift space - we include the analysis for (2+1)D projected distances as well.  The distances are now projected along the xy-plane and the redshift separation is calculated by using the coordinates and velocities of the galaxies in the z-direction, as well as the Hubble constant.  The right panel of Figure \ref{fig:roc_summary} shows the ROC curves of a classification algorithm using the projected distance in 2D and the redshift along the z-direction.  Similarly to the 3D analysis, we add in a domain alignment to account for different stellar formation models.  We find that the (2+1)D results do not differ much from the 3D ones, as the AUC and the purity-completeness values are very similar (see Figure \ref{fig:2d_thresh}).  From this, we conclude that most classification errors likely do not come from projection effects, as going from 3D to (2+1)D does not change the prediction accuracy significantly.

\begin{figure*}[ht]
    \includegraphics[width=0.95\linewidth]{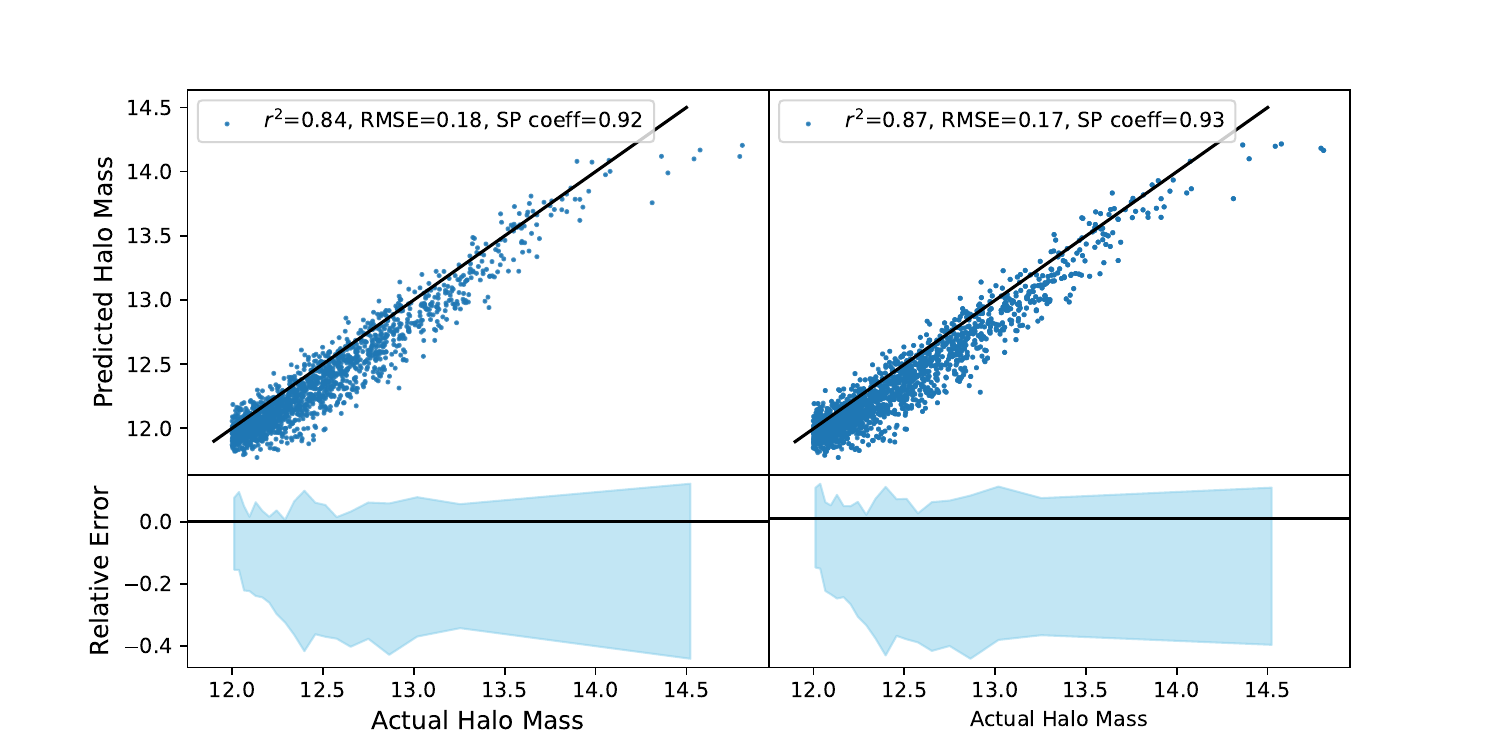}
    \centering
    \caption{TNG-trained halo-mass regressors evaluated on true EAGLE centrals in the (2+1)D representation after stellar-mass CDF alignment.  The left panel uses the fiducial neighbour threshold; the right combines the three thresholds. }
    \label{fig:2d_halos}
\end{figure*}

\begin{figure*}
    \centering
    \includegraphics[width=0.9\linewidth]{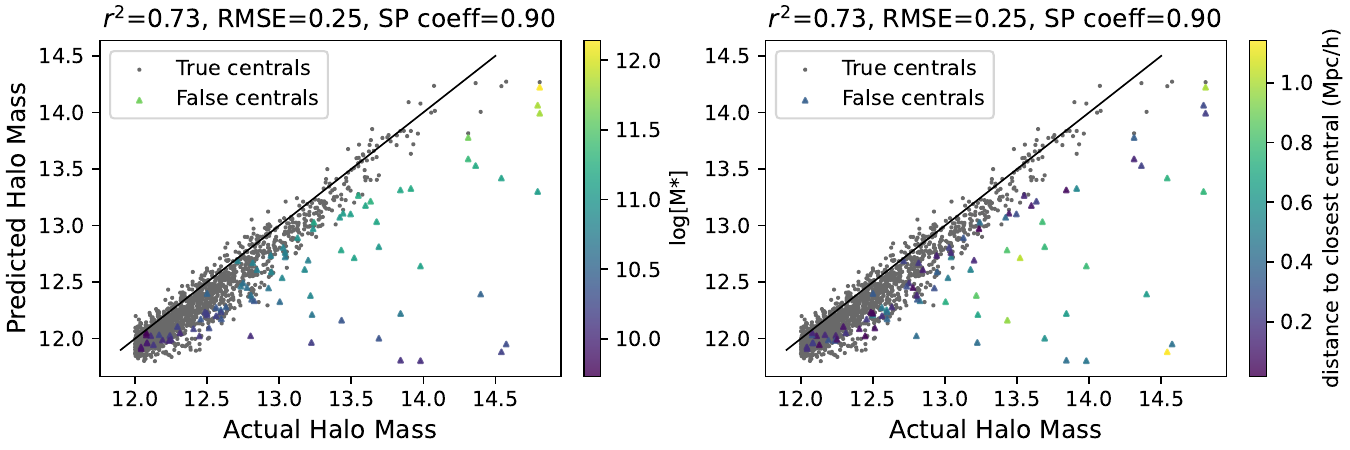}
        \caption{Fiducial mass cut of $10^9 M_{\odot}+$ of the predicted centrals, used to predict their host halos in (2+1)D, with the incorrectly identified centrals (actually satellites) coloured according to their stellar mass ratio and distance (in (2+1)D) to its closest central.  Again, the closest central is not necessarily the central of the satellite's host.  The most severe underpredictions are low stellar mass satellites misclassified as centrals.}
    \label{fig:2d_guess_predict}
\end{figure*}

We further include the host halo mass estimation for the (2+1)D projected distances, with the results shown in Figure \ref{fig:2d_halos}.  From Figures \ref{fig:roc_summary} and \ref{fig:2d_halos}, it can be seen that classification and host halo mass prediction work similarly well in the projection space as in the 3D space.  Similarly to the 3D case, domain-aligning the stellar mass improves results in the (2+1)D classification - see Table \ref{tab:mass_metrics}.  In Figure \ref{fig:2d_guess_predict}, it can be seen that the worst predictions come from misclassified centrals, again consistent with the 3D distance results.

\subsubsection{Summary of results on host halo mass predictions}

It can be seen here that having the same stellar mass distribution improves results - testing and training TNG yields higher $r^2$ values than testing on EAGLE.  Even with EAGLE, results improve with a stellar mass multiplier factor of 0.91 from the KS test for their stellar mass distributions.  Quantile transforming still does not yield an improvement in results.  Generally, the host halo mass can be predicted well, $r^2>0.7$ even in the worst case, with the simplest case of data wrangling (only adding a multiplicative factor to account for differences in stellar mass distributions).

\begin{table*}[p]
    \centering
    \begin{tabular}{|c|c|c|c|c|c|c|}
    \hline 
   Distance & Centrals & Neighbour selection & $M_{\ast}$ CDF aligned? & $r^2$ & RMSE [dex] & $\rho_{\rm S}$ \\  \hline    
    3D & Known & One $M_{\ast}$-cut & No & 0.81 & 0.19 & 0.92 \\  \hline
    3D & Known & One $M_{\ast}$-cut & Yes & 0.86 & 0.16 & 0.92  \\  \hline
    3D & Known & Different $M_{\ast}$-cuts & No & 0.85 & 0.19 & 0.94 \\  \hline
    3D & Known & Different $M_{\ast}$-cuts & Yes & 0.91 & 0.15 & 0.93  \\  \hline
    3D & Guessed & One $M_{\ast}$-cut & No & 0.69 & 0.26 & 0.90  \\  \hline
    3D & Guessed & One $M_{\ast}$-cut & Yes & 0.73 & 0.22 & 0.90  \\  \hline
    3D & Guessed & Different $M_{\ast}$-cuts & No & 0.76 & 0.26 & 0.93  \\  \hline
    3D & Guessed & Different $M_{\ast}$-cuts & Yes & 0.83 & 0.22 & 0.92 \\  \hline
    (2+1)D & Known & One $M_{\ast}$-cuts & No & 0.81 & 0.19 & 0.92 \\  \hline
    (2+1)D & Known & One $M_{\ast}$-cuts & Yes & 0.84 & 0.18 & 0.92 \\  \hline
    (2+1)D & Known & Different $M_{\ast}$-cuts & No & 0.84 & 0.19 & 0.94  \\  \hline
    (2+1)D & Known & Different $M_{\ast}$-cuts & Yes & 0.87 & 0.17 & 0.93  \\  \hline
    (2+1)D & Guessed & One $M_{\ast}$-cuts & No & 0.69 & 0.26 & 0.90  \\  \hline
    (2+1)D & Guessed & One $M_{\ast}$-cuts & Yes & 0.73 & 0.24 & 0.90  \\  \hline
    (2+1)D & Guessed & Different $M_{\ast}$-cuts & No & 0.74 & 0.26 & 0.92  \\  \hline
    (2+1)D & Guessed & Different $M_{\ast}$-cuts & Yes & 0.77 & 0.25 & 0.92  \\  \hline
    - & Known & Power law fit $\rm{M_{\ast}}$ to $\rm{M_{halo}}$ & Yes & \textbf{0.78} & \textbf{0.21} & \textbf{0.86} \\  \hline 
    - & Known & \textbf{$\rm{M_{\ast}}$ \xgb } & Yes & \textbf{0.80} & \textbf{0.20} & \textbf{0.86} \\  \hline 
    - & Guessed & Power law fit $\rm{M_{\ast}}$ to $\rm{M_{halo}}$ & Yes & \textbf{0.68} & \textbf{0.27} & \textbf{0.83} \\  \hline 
    - & Guessed & \textbf{$\rm{M_{\ast}}$ \xgb } & Yes & \textbf{0.69} & \textbf{0.26} & \textbf{0.84} \\  \hline 
    \end{tabular}
    \caption{\textbf{Cross-Dataset Host Halo Mass Inference}: Summary comparison of main results for host halo mass predictions on EAGLE data.  Like Table \ref{tab:tng_mass_metrics}, there are different metrics for the host halo mass regression task with different \xgb\ models, each trained and evaluated with different input features from TNG.  The model is evaluated on EAGLE galaxies that are either ``known" true centrals, or predicted centrals from a model trained on TNG data with neighbours selected with different $M_*$ cuts.  Features include distances to 10 nearest neighbours and the stellar mass of the central (either known true central or predicted central) and stellar masses of the neighbours.  The last four rows show the baseline prediction using a stellar-to-halo mass prediction, by both fitting a power law and using XGBoost on stellar mass only.  The top two rows of the last four are the stellar-to-halo mass for known centrals, while the last two rows are for guessed centrals.  The ``known" rows for the $M_{*}$ scaling only should be compared to 10th and 12th rows of this table, while the ``guessed" rows for the $M_{*}$ scaling only should be compared to 14th and 16th rows.}
    \label{tab:mass_metrics}
\end{table*}

\section{Discussion}\label{sec:discussion}

The galaxy classification is generally quite accurate, given the high AUC values across all training/testing samples, even when switching from one simulation, TNG, to another, EAGLE.  The central galaxy's neighbouring distances + stellar masses can also reliably predict most of their host halo masses, as from Table \ref{tab:mass_metrics}, it can be seen that the $r^2$ values are quite high, even when the training and testing are done on two different simulations and the testing sample consists of predicted centrals ($r^2=0.83$ being the best prediction). This shows that there is a strong correlation between the closest satellite-central distances + stellar mass and galaxy type, as well as the central galaxy's host halo mass.  Quantile transforming the data does not yield an improvement in results (see Table \ref{tab:scores}).  Instead, it was the inclusion of both neighbouring distances and stellar masses that yielded an improvement in results, showing that there is a necessity to go beyond the standard stellar-to-halo mass relation that has been used to predict halo mass in the past \citep{1976ApJ...203..297S, 2016MNRAS.459.2150W}.  Furthermore, multiplying the stellar mass by a domain alignment factor additionally improves results (see relative improvement for entries where ``$M_*$ CDF aligned?" is ``yes" in Table \ref{tab:mass_metrics} vs. entries where this is ``no").  This domain alignment accounts for TNG and EAGLE having different star formation prescriptions that lead to different stellar mass functions and stellar-to-halo mass relations.  We should also note that the AUC values from the 3D distance classification did not change significantly when switching to (2+1)D projected distances (see Figure \ref{fig:3d_roc} and \ref{fig:2d_roc}).  From this, it seems that projection effects do not change the galaxy classification results in any significant way.

Our analysis using both neighbouring stellar masses and distances results in a better prediction than the standard stellar-to-halo mass relation, which is often used to estimate halo masses \citep{Girelli:2020goz}.  The last four rows of Table \ref{tab:mass_metrics} show the prediction metrics (in bold) of the host halo mass from using both a power law scaling relation with the $M_{\ast}$ as well as by inputting $M_{\ast}$ into \xgb.  Both our 3D and (2+1)D predictions fared better than the stellar-to-halo mass estimations, as did both the guessed and known centrals.  Note that we also CDF align the stellar mass function for the cross-dataset inference using a stellar-to-halo mass relation.

We also see that when combining both galaxy classification and host halo mass prediction - predicting the host mass of the predicted centrals (see Figures \ref{fig:3d_guess_predict} and \ref{fig:2d_guess_predict}) - the largest errors are underpredictions, driven by a combination of lower stellar masses of the misclassified ``central" (actual satellite) and the misclassified ``central" being closer to a lower mass halo than the central of its own host.  The few misclassified galaxies that are closer to their own centrals than another halo's central - but are still significantly underpredicting their host halo mass - have substantially less stellar mass than the true central of their host halo, which is what likely drives this host mass underprediction.
In summary, the largest errors come from incorrectly classified galaxies either in their host halos' outskirts - which happen to be closer to halo environments with smaller masses than those of their own host halos - and/or have low stellar masses in comparison to their true host centrals, see Figures \ref{fig:3d_incorrect} and \ref{fig:2d_incorrect}.

\subsection{Comparison to previous works}
There are several approaches in the literature to address the problem of central/satellite galaxy classification and host halo estimation based on environmental measures. \citet{Oguri:2014dqa} is one of the earlier works to utilize a counts-in-cylinder approach to characterize dark matter halo masses of galaxies, assuming an HOD framework for central/satellite classification to define a probability distribution function for their host halo mass inference task.  We have included an additional component of explicitly inferring central/satellite classification as a step in our inference pipeline, which can directly yield a halo catalogue with estimated halo masses for observed galaxies that have measurements of right ascension (RA), declination (DEC), and either spectroscopic or photometric redshifts.  Furthermore, instead of counting neighbours, we use the spatial distribution of galaxies around the central to classify the host halo mass.

In \citet{2023OJAp....6E..37B}, they compare how counts-in-cylinder information and distances to the k-nearest neighbours (explicitly utilized in our analysis) inform dark matter halo mass estimation, finding these to provide similar prediction power with neural networks.  Our method of host halo mass estimation with an \xgb\ model that utilizes neighbour information from known centrals has a similar RMSE to that found in \citet{2023OJAp....6E..37B}.  They further do a similar central/satellite classification task in a follow-up work \citep{Bowden:2025djs}, where they find comparable results to ours when looking at correctly classified centrals in the same stellar mass range.  However, their results were obtained from training and testing on the same Universe Machine \citep{2019MNRAS.488.3143B} output data.  In our analysis, we trained on data from magnetohydrodynamical simulations \citep{Nelson:2018uso, Vogelsberger:2019ynw} from TNG100, then tested on another similar simulation that the training had no information on.

Finally, in \citet{Calderon:2019ofx}, there exists a complementary study to constrain dark matter halo masses from galaxy properties using \xgb\ and other machine learning algorithms, finding that these better constrained masses in out-of-distribution test datasets than traditional abundance matching and dynamical methods. With a simplified feature set (only neighbour distances and stellar masses), we have identified a potential mitigation strategy for evaluating an \xgb\ halo mass estimator on an out-of-distribution dataset by adding a correction factor to align the CDFs of the stellar masses in the dataset. Our errors in halo mass estimation are comparable to those found in the work presented in \citet{Calderon:2019ofx}; there is an average of around $20\%$ scatter in the deviation from the host halo mass.

\subsection{Limitations and Future Prospects}
In this work, we used 10 distances as an arbitrary limit, although we found that generally 5-7 distances was enough for our results to be consistent.  It would be interesting to test the optimal number of distances needed to correctly predict the host halo mass and if they differ for different host halo masses - between $10^{12}-10^{14}$ - and different simulations.  We also only used central galaxies to predict host halo masses here; no work was done about predicting subhalo masses from satellite distribution.  Because satellite galaxies can have a much wider stellar mass/colour/spatial distribution compared to central galaxies due to their generally more complex environmental history, it is possible that subhalo masses would require more parameters than just spatial distribution and stellar mass to predict, and likely more than 10 neighbouring distances as well.  The optimal number of neighbours, as well as the subhalo mass predictions, are beyond the current scope of this work.

In this work, we do not explicitly distinguish between the one-halo and two-halo regimes when considering galaxy environments, i.e., do not specify whether the 10 closest neighbours are part of the same halo as the central galaxy.  Instead, we allowed XGBoost to learn the relevant relationships directly from the available neighbouring stellar masses and spatial information, mimicking the lack of information about halo membership in real observations.  An important test of this method's applicability to observations will be to determine how its performance changes when using projected galaxy separations, line-of-sight information, and stellar masses inferred from luminosity, as opposed to explicitly given simulated information.  The real test of our proposal would then be to apply our methodology and analyses to observational galaxy catalogues.  We should then account for differences between the stellar mass distributions of simulations and observations by using a stellar mass correction factor analogous to what we used for TNG100 vs. EAGLE above.  One potential intermediate follow-up step can be to first test the method on simulated lightcones, prior to examining real space observational catalogues such as SDSS \citep{2026ApJS..285....9S}, DES \citep{DES:2021wwk}, LSST \citep{2020arXiv200111067B}, and Euclid \citep{Euclid:2024yrr}.

From Figures \ref{fig:3d_guess_predict2} and \ref{fig:2d_guess_predict2}, it can be seen the worst underpredictions of host halo mass are associated with misclassified satellites physically closer to a lower mass halo's central than its own host's central.  We suspect these galaxies could be from an actively merging system, either on its first infall or splashback.  The local environment of these misclassified galaxies reflects the lower mass system, but has been identified by the algorithm as being associated with the more massive system.  While these misclassified galaxies are problematic in our analysis, they could be potentially exploited to find ongoing mergers and splashback galaxies \citep{10.1111/j.1365-2966.2004.08562.x, 10.1093/mnras/stad045}.  By combining their positions on the galaxy outskirts with their spatial distribution, stellar mass, and velocity profiles in a future work, it could be possible to find a method to classify galaxies with $R_{vir} \leq R_{host} \leq 2R_{vir}$ \citep{10.1111/j.1365-2966.2004.08562.x} as splashback galaxies.  The combination of a misclassified satellite both being closer to a neighbouring halo, and having a spatial distribution + stellar mass similar to that of a central is an interesting case to study, and could potentially be connected to infalling or splashback galaxies.

Finally, simulations have a limited box size, which in this case was $(75 Mpc/h)^3$.  As a result, we do not have halos larger than $\sim 10^{14.5}M_{\odot}$, and a limited number of cluster sized halos in general.  However, in observations, $10^{15}+M_{\odot}$ halos have been observed, so whether this analysis can be extrapolated to them the same way will be an interesting test for the future - possibly trained and tested on COLIBRE \citep{Schaye:2025xuv} as a proxy for observations.  We also impose mass cuts on halos and stellar mass cuts on galaxies to take into account resolution limits - how our analysis will change for irregular dwarves with halos $\leq 10^{11} M_{\odot}$ is still an open question for exploration.

\section{Conclusion}\label{sec:conclusion}
In this work, we have shown that using the stellar mass \footnote{in observations, photometric or spectroscopic redshifts} and spatial distribution (via neighbouring distances) of galaxies with the XGBoost algorithm is enough to classify the galaxy type as central or satellite and predict the host halo mass using the central properties.  Quantile transforming the training properties do not yield an improvement in the results, but including a domain alignment of 0.91 did, so we used the original properties with a multiplication factor of 0.91.  Switching from real space to redshift space projected distance also does not yield a significant difference in results, as the AUC and purity-completeness threshold values stay relatively constant.  

When it comes to the FOF halo mass of the galaxies' host, similar results apply - quantile transformation again does not improve the results, while adding a domain alignment to account for differences in stellar mass distributions yields significant improvement here.  The largest errors from the mass prediction comes from misclassified central galaxies - if the galaxy type is known with $100\%$ certainty, then the estimated vs. actual host halo mass consistently has a $r^2 > 0.8$, even above 0.9 if the same simulation is used in the testing and training sample.  The largest discrepancies between predicted and actual halo mass occur when the misclassified galaxy in question is closer to another less massive halo's central than its own central, and/or has a significantly lower stellar mass than its central.  The misclassified galaxy is likely associated with environmental features from the smaller neighbouring halo.  In future work, we can investigate potential mitigation methods, as well as the relationship between these misclassified galaxies to the merging history of its host halo.

Here, we have shown that using only spatial distribution and the stellar masses of galaxies is enough to predict both galaxy type and host halo mass.  This methodology can further be extended to infer halo concentration, halo membership, and create localized mass maps.  Future prospects include testing our analysis on observational data and optimizing for number of neighbours used, as well as extending this work to predict subhalo masses.

\section*{Acknowledgments}
AC and NA are partially supported by the Natural Sciences and Engineering Research Council of Canada and Perimeter Institute for Theoretical Physics; research at the Perimeter Institute is supported in part by the Government of Canada through the Department of Innovation, Science and Economic Development Canada and by the Province of Ontario through the Ministry of Colleges and Universities. 
CA acknowledges support from the Leinweber Institute, DOE grant DE-SC0019193.  This research was supported in part by the Simons Foundation through the Simons Foundation Emmy Noether Fellows Program at Perimeter Institute. YTL acknowledges support from the National Science and Technology Council grant NSTC 115-2112-M-001-018.

\appendix

\section{Appendix: Comparison of Different Classification Methods for Galaxies}

The results for all our analysis above were obtained using XGBoost. However, before going to that, we tested some other classification methods as well (both analytical/statistical and other machine learning algorithms).  The summarized results of those are outlined below.

\subsection{Other Machine Learning Algorithms}

We perform a brief test of model performance comparisons on the central vs. satellite classification task with Logistic Regression, a simple two-layer fully connected neural network, then XGBoost.  The input features are still neighbouring distances and stellar masses.  We use both the original simulation data and also features that have been quantile transformed in the analysis.  Quantile transforming neighbour distances as input features improves the performance for Logistic Regression and the 2-layer neural network, while quantile transformed features do not improve performance with XGBoost.  In either case, XGBoost had the best performance overall (see Table \ref{tab:scores}), which is why we ended up going with that for our final analysis.

When including the stellar mass of the neighbouring galaxies along with the distances, the value of $r^2$ increases.  When including other galaxy properties as features, quantile transforming the features makes little difference in model performance. Note that the addition of the galaxy's own internal properties beyond stellar mass have a negligible change on the model's performance. 

\begin{table*}[htp!]
    \centering
    \begin{tabular}{|c|c|c|c|c|}
    \hline
    Method & Quantile Transformed? & Distance only AUC & Distances + $M_*$ AUC \\ 
    \hline
    Logistic Regression & No & 0.83 & 0.91 \\ \hline
    Logistic Regression & Yes & 0.97 & 0.97 \\ \hline
    Neural Nets & No & 0.5 & 0.90 \\ \hline
    Neural Nets & Yes & 0.99 & 0.90 \\ \hline
    XGBoost & No & 0.93 & 1.0 \\ \hline
    XGBoost & Yes & 0.99 & 1.0 \\ \hline
    \end{tabular}
    \caption{AUC values for the ROC curves corresponding to different models trained with different features to predict central vs. satellite classification. }
    \label{tab:scores}
\end{table*}

\subsection{Predicted centrals error summary}
\begin{figure*}[ht]
    \begin{subfigure}[t]{0.48\linewidth}
        \centering        \includegraphics[width=0.95\linewidth]{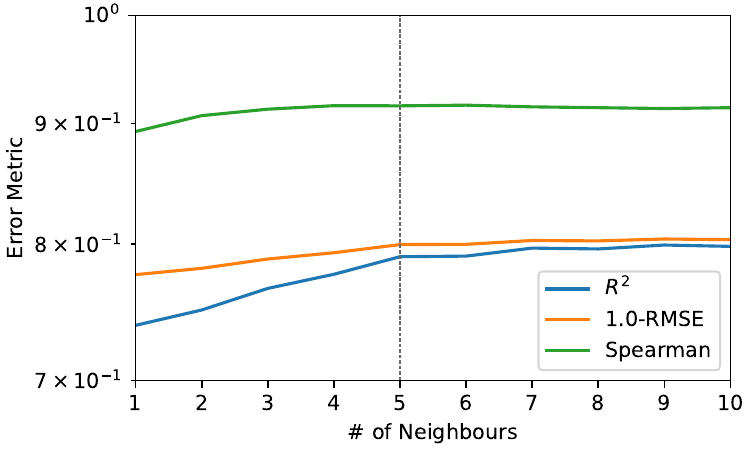}
        \caption{Performance metrics found using (2+1)D projected distances.}
    \end{subfigure}
    \begin{subfigure}[t]{0.48\linewidth}
        \centering        \includegraphics[width=0.95\linewidth]{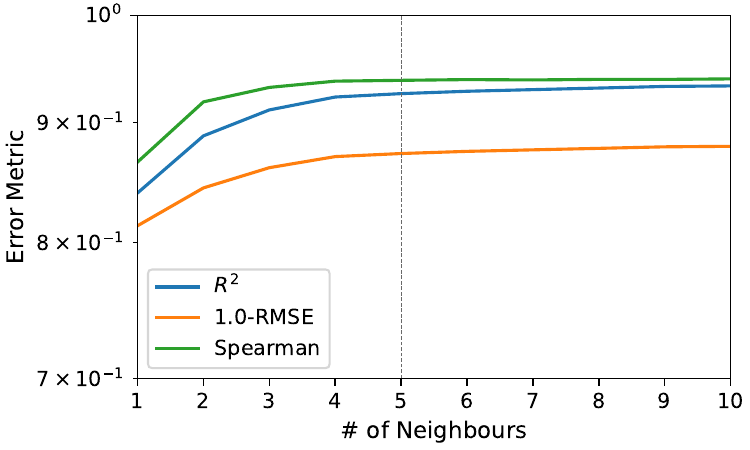}
        \caption{Performance metrics found using 3D distances.}
    \end{subfigure}
    \centering
    \caption{Summary of the halo mass estimation performance metrics using each galaxy's neighbouring distance, up to the 10 closest distances. }
    \label{fig:distance metrics}
\end{figure*}

We used the closest 10 neighbours for a more comprehensive study, but not all 10 nearest neighbours are actually needed for good halo mass estimation results in Sec. \ref{sec:results_tng_mass} - this is shown in Figure \ref{fig:distance metrics} where we see the errors plateau around the fifth closest neighbour.

\begin{figure*}
    \includegraphics[width=0.95\linewidth]{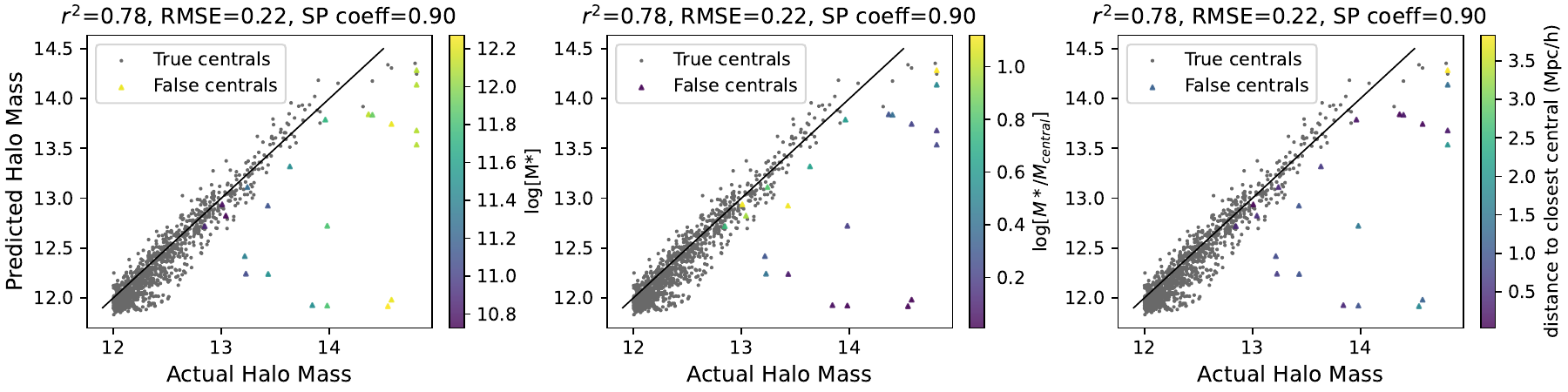}
        \caption{Fiducial mass cut of $10^9 M_{\odot}+$ of the predicted centrals, used to predict their host halos, with the incorrectly identified centrals (actually satellites) coloured according to their stellar mass ratio and distance to its actual central, which is usually, but not always, the central closest to the satellite.  Using full 3D distances.}
    \label{fig:3d_guess_predict2}
\end{figure*}

\begin{figure*}
    \includegraphics[width=0.95\linewidth]{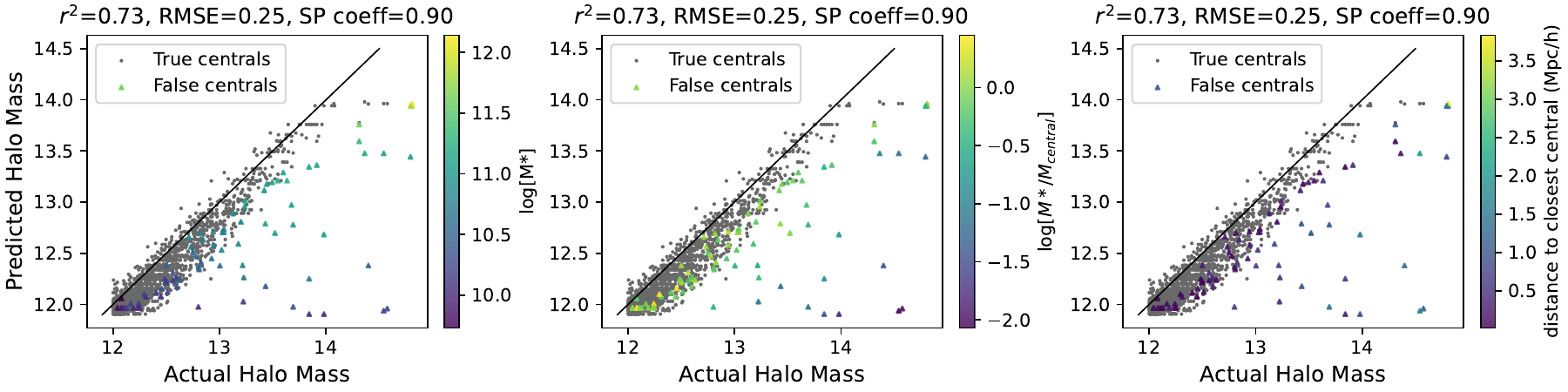}
        \caption{Fiducial mass cut of $10^9 M_{\odot}+$ of the predicted centrals, used to predict their host halos, with the incorrectly identified centrals (actually satellites) coloured according to their stellar mass ratio and distance to its actual central, which is usually, but not always, the central closest to the satellite.  This is in (2+1)D redshift space distances.}
    \label{fig:2d_guess_predict2}
\end{figure*}

Finally, we show that the 3D full distances do a slightly better job of estimating the host halo mass of the galaxies from the $r^2$ and RMSE values in Figures \ref{fig:3d_guess_predict2} and \ref{fig:2d_guess_predict2}.  We also see that the largest errors in predicted mass come from misclassified galaxies; both results are consistent with Figures \ref{fig:3d_guess_predict} and \ref{fig:2d_guess_predict} in Sec. \ref{sec:results} above.  We further include the distance and stellar mass ratios between the misclassified galaxies and their actual centrals here to further emphasise that satellites with either large stellar masses or small distances to a central are more likely to be misclassified.

\clearpage
\bibliographystyle{aasjournal}
\bibliography{references}

\end{document}